# Light dynamics in glass-vanadium dioxide nanocomposite waveguides with thermal nonlinearity

Yaroslav V. Kartashov[1] and Victor A. Vysloukh[2]

[1]*ICFO-Institut de Ciencies Fotoniques, and Universitat Politecnica de Catalunya, Mediterranean Technology Park, 08860 Castelldefels (Barcelona), Spain*

[2]*Departamento de Fisica y Matematicas, Universidad de las Americas – Puebla, Santa Catarina Martir, 72820, Puebla, Mexico*

We address the propagation of laser beams in $SiO_2 - VO_2$ nanocomposite waveguides with thermo-optical nonlinearity. We show that the large modifications of the absorption coefficient as well as notable changes of refractive index of $VO_2$ nanoparticles embedded into the $SiO_2$ host media that accompany the semiconductor-to-metal phase transition may lead to optical limiting in the near-infrared wave range.

The importance of thermal self-action of laser beams was realized almost forty years ago [1,2]. In some materials, such as $VO_2$, heating results in abrupt changes of absorption and refraction [3,4] due to semiconductor-to-metal phase transition at $T_{\mathrm{pt}} \approx 67\,^{\mathrm{o}}\mathrm{C}$. In $VO_2$ thin films such transitions can be driven optically at subpicosecond time scales [5,6]. Optical properties of $VO_2$ nanocomposites depend on sizes of the embedded particles [7-9], allowing various switching applications [7]. Such composites and films possess ultrafast nonlinearities [10] and enhanced absorption [11,12]. Phase transition is affected by the film morphology [13]. It can be used to achieve switchable reflectivity of nanocomposite layers [14]. However, for quasi-cw illumination the properties of $VO_2$ films [15] and nanoporous glass-$VO_2$ composites [16-18] are altered mostly by slow phase transition due to light heating: a mechanism that is completely different from the ultrafast optically induced phase transition. Moreover, in thick nanocomposites with low concentration of doping nanoparticles ($\nu \sim 10^{-2}$ %) light propagation is affected not only by thermally induced changes of absorption, but also by the corresponding refractive index variations [16,17]. Thus, exploration of laser beam dynamics in the waveguiding geometries is rather important. In this Letter we study light dynamics in $SiO_2-VO_2$ nanocomposite waveguides for thermally-induced semiconductor-to-metal phase transition. Optical limiting in this setting can be controlled by the initial system temperature and by the intensity of the input light beam.

Light propagation in a planar waveguide formed by $SiO_2$ glass cladding and a glass core with embedded spherical $VO_2$ nanoparticles is described by the nonlinear Schrödinger equation for the dimensionless field amplitude $q(\eta,\xi)$:

$$i\frac{\partial q}{\partial \xi}=\frac{1}{2}\frac{\partial^2 q}{\partial \eta^2}+\{\,p_{\mathrm{r}}[1+\mu_{\mathrm{r}}S(T)]-ip_{\mathrm{i}}[1+\mu_{\mathrm{i}}S(T)]\}\,R(\eta)q, \tag{1}$$

where $\xi$ is the propagation distance normalized to $L_{\mathrm{dif}}=kx_0^2$, $k=2\pi n_{\mathrm{rh}}/\lambda$, $n_{\mathrm{rh}}$ is the real part of the refractive index of host material, $\lambda$ is the wavelength. The transverse coordinate $\eta$ is expressed in units of core width $x_0=10\ \mu\mathrm{m}$, while the function $R(\eta)=\exp(-\eta^8)$ describes concentration profile of $VO_2$ nanoparticles. The parameter $p_{\mathrm{r}}=2\pi L_{\mathrm{dif}}(n_{\mathrm{rs}}-n_{\mathrm{rh}})/\lambda$ is the normalized difference of wavenumbers in the core (semiconducting phase is assumed) and glass cladding; $\mu_{\mathrm{r}}=(n_{\mathrm{rm}}-n_{\mathrm{rs}})/(n_{\mathrm{rs}}-n_{\mathrm{rh}})$ is the normalized difference of refractive indices in metallic $n_{\mathrm{rm}}$ and semiconducting $n_{\mathrm{rs}}$ phases; $p_{\mathrm{i}}=2\pi L_{\mathrm{dif}}(n_{\mathrm{is}}-n_{\mathrm{ih}})/\lambda$ characterizes the difference of imaginary parts of refractive indices, while $\mu_{\mathrm{i}}=(n_{\mathrm{im}}-n_{\mathrm{is}})/(n_{\mathrm{is}}-n_{\mathrm{ih}})$. The smoothed step-like function $S(T)=\{1+\tanh[(T-T_{\mathrm{pt}})/\delta T]\}/2$ of the temperature $T$ describes the phase transition. The 10-90% width of the semiconductor-to-metal transition curve at high temperature $W$ is related with parameter $\delta T$ by $W\simeq 2.2\delta T$ (which amounts to $\sim 11\ ^\circ\mathrm{C}$ in nanoporous glass-$VO_2$ composites [18]). We used Maxwell Garnett formula $\varepsilon_{\mathrm{c}}=\varepsilon_{\mathrm{h}}+\nu\varepsilon_{\mathrm{h}}(\varepsilon_{\mathrm{s}}-\varepsilon_{\mathrm{h}})/[\varepsilon_{\mathrm{h}}+(1-\nu)(\varepsilon_{\mathrm{s}}-\varepsilon_{\mathrm{h}})/3]$ to calculate the complex dielectric constant $\varepsilon_{\mathrm{c}}=(n_{\mathrm{rs}}-in_{\mathrm{is}})^2$ of composite material with volume concentration $\nu$ of $VO_2$ nanoparticles. We suppose that the particle diameter (10-20 nm) is small enough to neglect scattering in comparison with absorption that allows one to use an effective media approximation [16]. The profile of normalized temperature $\theta=(T-T_0)/\delta T$ in planar geometry is described by the equation:

$$\frac{\partial \theta}{\partial \tau}-\frac{\partial^2 \theta}{\partial \eta^2}=p_{\mathrm{nl}}[1+\mu_{\mathrm{i}}S(\theta-\theta_0)]R(\eta)|q|^2, \tag{2}$$

where the time $\tau$ is normalized by $\tau_0 = x_0^2 / \kappa$ ; $\kappa = \chi / \rho C$ is the thermo-diffusion coefficient; $\theta_0 = (T_{\rm pt} - T_0) / \delta T$ is the normalized difference between phase transition temperature and the ambient one $T_0 < T_{\rm pt}$ ; $p_{\rm nl} = I_0 / I_{\rm n}$ is defined by the peak laser beam intensity $I_0$ normalized by $I_{\rm n} = \rho C \lambda \delta T / (4\pi n_{\rm is} \tau_0)$ . The edges of planar waveguide at $\eta = \pm L_{\rm b} / 2$ are thermo-stabilized at $T = T_0$ , while upper and lower facets are thermo-isolated, so along one transverse coordinate the intensity distribution is uniform and temperature is constant. The system of Eqs. (1,2) was solved for input beam $q(\eta, \xi = 0) = \exp(-\eta / \eta_0^2)$ , whose width $\eta_0 = 1.38$ was selected to match the width of linear mode of lossless waveguide at $p_{\rm r} = 1$ .

Importantly, the size of embedded nanoparticles influences the width of hysteretic loop and the temperature of phase transition [8,19], as well as optical constants of composite material. Since experimental dependencies of complex dielectric constants on $\lambda$ are not currently available, we used optical constants of $VO_2$ films [4], as well as results of experiments with nanoporous glasses [18] to estimate the range of variation of parameters. The data of simulations can be rescaled using normalizations given above to any specific width of semiconductor-to-metal transition curve for the particular setting and size of nanoparticles. It should be stressed that the very possibility of a phase transition in different nanocomposite samples with 3-40 nm particles was proven experimentally in [18,19].

The modifications in the refractive index of $VO_2$ nanoparticles essentially depend on $\lambda$ . Thus, at $\lambda \simeq 1\ \mu$m upon the semiconductor-to-metal phase transition the absorption coefficient grows more then six times, while refractive index of guiding core drops off by $\sim 20\%$ at $\nu = 0.02\%$ (Fig. 1). At $\lambda \simeq 1.5\ \mu$m the increase of absorption is even more considerable. Due to this remarkable absorption growth, the optical limiting is possible in the near-infrared wavelength range. Initially relatively low laser beam absorption produces the temperature growth that speeds up absorption (due to phase transition) and finally blocks the guided light. This process is shown in Fig. 2(a) where the light intensity distributions in the $(\eta, \xi)$ plane are depicted for different time moments. Thermally-induced diminishing of the refractive index produces broadening of guided beam. At the wavelength $\lambda \simeq 1.5\ \mu$m the initial absorption in semiconductor state is smaller which results in slower switching [Fig. 2(b)]. Scaling factors at wavelengths $\lambda \simeq 1.0\ \mu$m and $\lambda \simeq 1.5\ \mu$m are summarized in Table I.

The characteristic features of optical limiting are illustrated by Fig. 3 at $\lambda \simeq 1.5\ \mu$m . The growth of the maximal input temperature $\theta_{\rm in}(\tau) = \theta(\eta = 0, \xi = 0, \tau)$ versus time is shown in Fig. 3(a) in comparison with the output temperature $\theta_{\rm out}(\tau) = \theta(\eta = 0, \xi = L, \tau)$ (here $L = 2$ is the sample length along $\xi$ axis, $p_{\rm nl} = 1, \theta_0 = 2$). One can clearly see how heating speeds up near the point $\theta \simeq 2$ due to increase in absorption induced by phase transition. The temperature at $\eta = 0$ decreases along the waveguide [Fig. 3(b)] at any moment of time, while the border dividing the semiconducting and metallic phases of doping nanoparticles (i.e. the point where $\theta = 2$) gradually shifts into the waveguide depth, as indicated by an arrow. Figure 3(c) shows variation of light intensity along the waveguide axis for different moments of time. The difference of the fading rates in the beginning of the waveguide (metallic phase) and in its rear part (which still holds in semiconducting phase) is clearly visible. Figure 3(d) illustrates the switching characteristics (i.e., dependencies of the output peak intensity $I_{\rm out}(\tau) = |q(0, L, \tau)|^2$ on time) for various values of $\theta_0$ proportional to difference between the phase transition temperature and the ambient temperature. Notice that the high-contrast optical limiting is possible at $\theta_0 \simeq 2$. Another essential parameter for control of switching is the waveguide length $L$. Figure 3(e) shows the maximal output peak intensity $I_{\tau=0}$ in the very beginning of optical pulse (at $\tau = 0$) and minimal steady-state output intensity value $I_{\tau\to\infty}$ as functions of $L$. Importantly, the intensity $I_{\tau\to\infty}$ decreases with $L$ much faster than $I_{\tau=0}$ due to two factors: Rapid growth of absorption that accompanies the increase of temperature and de-trapping of optical radiation due to decrease of refractive index of guiding core. The mean transmission $I_{\rm m} = (I_{\tau=0} + I_{\tau\to\infty})/2$ diminishes, while switching contrast $V = (I_{\tau=0} - I_{\tau\to\infty})/(I_{\tau=0} + I_{\tau\to\infty})$ monotonically increases with $L$ as illustrated in Fig. 3(f). Switching/limiting control might be accomplished by absorption of a single optical pulse preferably with duration less then the thermo-diffusion time $\tau_0$. For instance, in 100 $\mu$m$^2$ mode area waveguide the energy of $\sim \mu$J carried by $\mu$s optical pulse would be sufficient to realize 70% switching contrast. At the same time optical transmission can also be effectively controlled by thermalized cw radiation or external heat sources/sinks.

In conclusion, we studied light propagation dynamics in $SiO_2 - VO_2$ nanocomposite waveguide during the semiconductor-to-metal phase transition. We showed that optical limiting contrast and transient time can be controlled by the waveguide length and by detuning of the initial waveguide temperature from that of the phase transition. Importantly, the waveguiding geometry offers unique opportunity to confine optical radiation in a small area of nanocomposite core and arrange long-path radiation-material interaction.

## References with titles

## References without titles

## Figure captions

Figure 1. Real (a) and imaginary (b) parts of refractive index of $SiO_2-VO_2$ nanocomposite in metal (black curves) and semiconductor (red curves) phases versus $\nu$ (in percents) at $\lambda = 1\ \mu$m.

Figure 2. (a) Optical limiting at $\lambda = 1.0\ \mu$m for $p_{\rm i} \simeq 0.264$, $\mu_{\rm i} \simeq 5.382$, $p_{\rm r} \simeq 1.323$, $\mu_{\rm r} \simeq -0.201$, and (b) at $\lambda = 1.5\ \mu$m for $p_{\rm i} \simeq 0.056$, $\mu_{\rm i} \simeq 7.928$, $p_{\rm r} \simeq 0.661$, $\mu_{\rm r} \simeq 0.747$. Spatial intensity distributions are shown in different moments of time. In all cases $\nu = 0.02\,\%$.

Figure 3. (a) Maximal input and output temperatures versus $\tau$. Temperature (b) and peak intensity (c) versus $\xi$. (d) Output peak intensity versus $\tau$. Circles in (a) and (b) correspond to points of phase transition. In (a)-(d) $L = 2$. Output peak intensities at $\tau = 0$ and $\tau \to \infty$ (e) and mean transmission and switching contrast (f) versus $L$. In all cases $\lambda = 1.5\ \mu$m and $\nu = 0.02\,\%$.

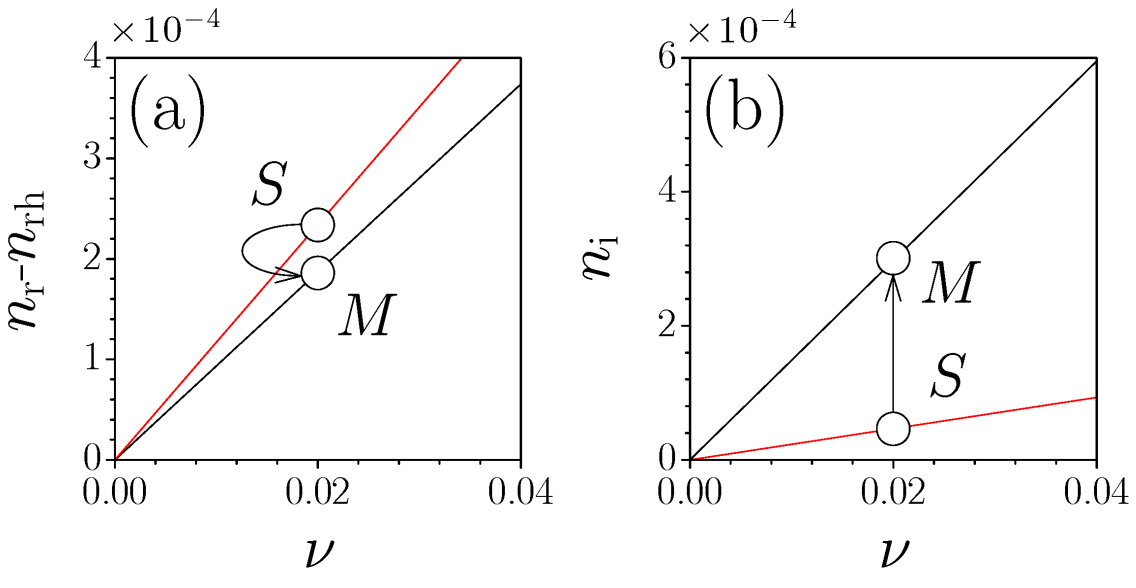

Figure 1. Real (a) and imaginary (b) parts of refractive index of $SiO_2 - VO_2$ nanocomposite in metal (black curves) and semiconductor (red curves) phases versus $\nu$ (in percents) at $\lambda = 1\ \mu$m.

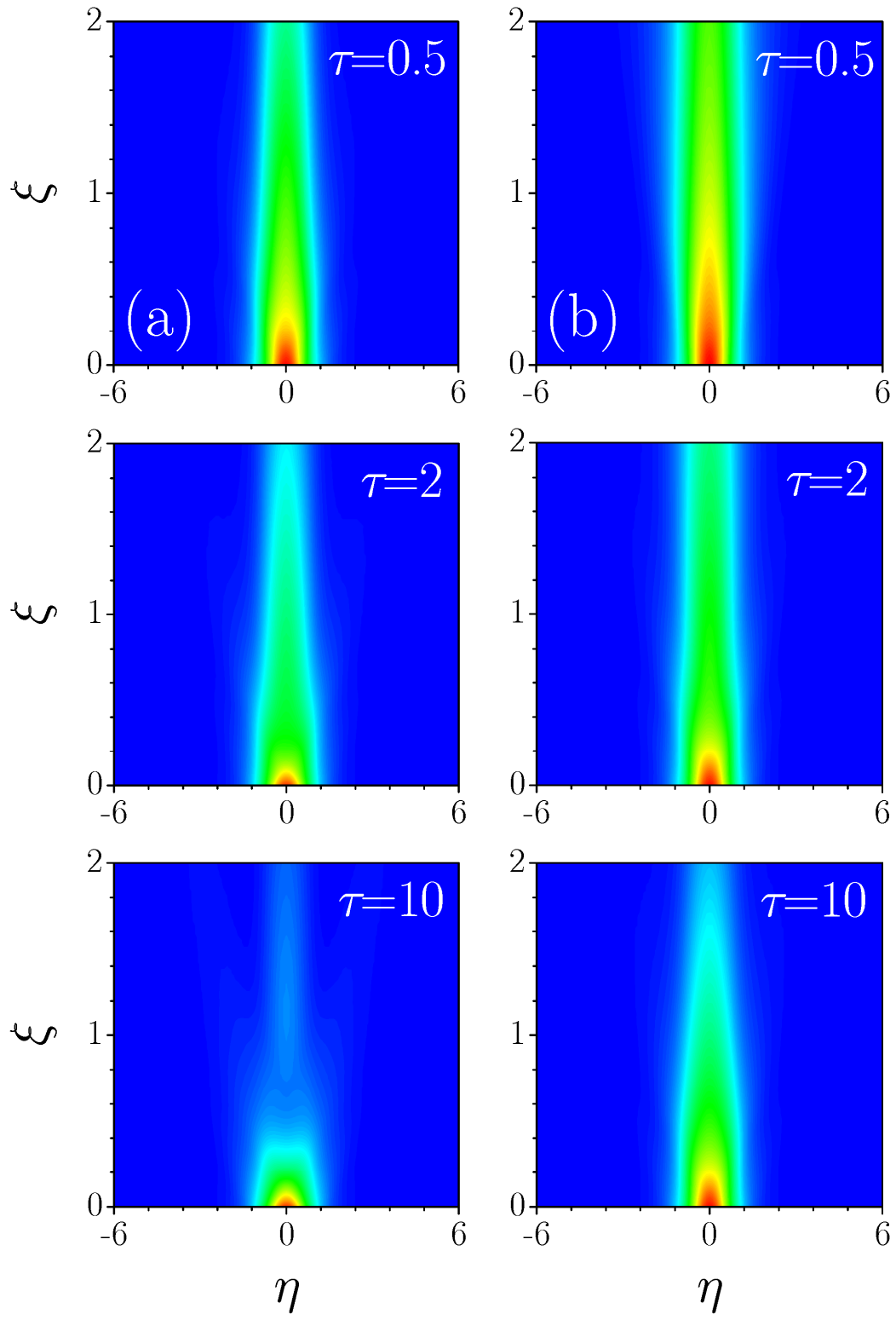

Figure 2. (a) Optical limiting at $\lambda = 1.0\ \mu$m for $p_{\rm i} \simeq 0.264$, $\mu_{\rm i} \simeq 5.382$, $p_{\rm r} \simeq 1.323$, $\mu_{\rm r} \simeq -0.201$, and (b) at $\lambda = 1.5\ \mu$m for $p_{\rm i} \simeq 0.056$, $\mu_{\rm i} \simeq 7.928$, $p_{\rm r} \simeq 0.661$, $\mu_{\rm r} \simeq 0.747$. Spatial intensity distributions are shown in different moments of time. In all cases $\nu = 0.02\,\%$.

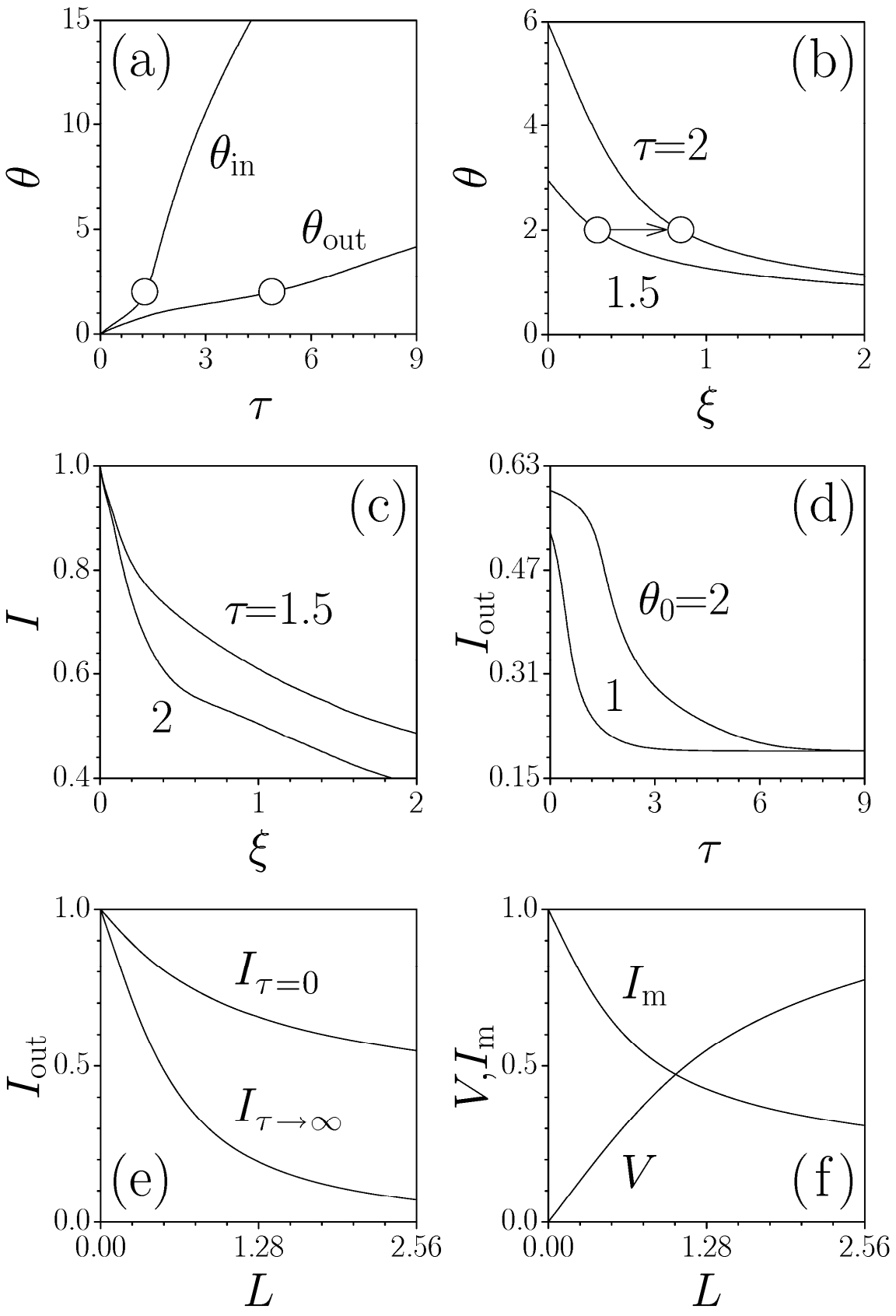

Figure 3. (a) Maximal input and output temperatures versus $\tau$ . Temperature (b) and peak intensity (c) versus $\xi$ . (d) Output peak intensity versus $\tau$ . Circles in (a) and (b) correspond to points of phase transition. In (a)-(d) $L = 2$ . Output peak intensities at $\tau = 0$ and $\tau \to \infty$ (e) and mean transmission and switching contrast (f) versus $L$ . In all cases $\lambda = 1.5\ \mu$m and $\nu = 0.02\,\%$ .

| $\lambda$, $\mu$m | $x_0$, $\mu$m | $L_{\mathrm{dif}}$, mm | $\tau_0$, $\mu$s | $I_{\mathrm{n}}$, kW/cm$^2$ |
|---|---|---|---|---|
| 1.0 | 10 | 0.91 | 0.12 | 11.6 |
| 1.5 | 10 | 0.61 | 0.12 | 36.1 |

Table I. Scaling factors at $\lambda \simeq 1.0\ \mu$m and $\lambda \simeq 1.5\ \mu$m.